\documentclass[]{aa} 
 
\usepackage{natbib}  

\usepackage{graphicx}
\usepackage[varg]{txfonts}
\usepackage{hyperref}

\def\Ha   {$\rm H\alpha$}
\def\Liu  {\ion{Li}{i}}
\def\Bed  {\ion{Be}{ii}}

\def\Nau  {\ion{Na}{i}}
\def\Mgu  {\ion{Mg}{i}}

\def\Alu  {\ion{Al}{i}}
\def\Siu  {\ion{Si}{i}}

\def\Cau  {\ion{Ca}{i}} 
\def\Cad  {\ion{Ca}{ii}} 
\def\Scd  {\ion{Sc}{ii}} 
\def\Tiu  {\ion{Ti}{i}}   
\def\Tid  {\ion{Ti}{ii}}
\def\Cru  {\ion{Cr}{i}}

\def\Feu  {\ion{Fe}{i}}
\def\Fed  {\ion{Fe}{ii}}
\def\Cou  {\ion{Co}{i}}
\def\Niu  {\ion{Ni}{i}}
\def\Srd  {\ion{Sr}{ii}}

\def\Teff  {$T_\mathrm{eff}$}
\def\logg  {$\log g$}

\def\vt    {$\rm v_{t}$}
\def\kms   {$\rm km\,s^{-1}$}
\def\JJ    {2MASS\,J1808-5104 }
\def\JJj   {2MASS\,J1808-5104}
\def\BD    {BD\,$+44^\circ 493$}

\newcommand{\cobold}{{CO$^5$BOLD}}

\begin{document}

\title{ 
Be and O in the ultra metal-poor dwarf 
2MASS\,J18082002-5104378: 
The Be-O correlation
\thanks{Based on observations collected at the
European Organisation for Astronomical Research in the Southern
Hemisphere under ESO programmes 101.A-0229(A), (PI M.Spite) and
293.D-5036 (PI J.M{\'e}lendez). This research has also made use of Keck Observatory Archive (KOA), operated by the W. M. Keck Observatory and the NASA Exoplanet Science Institute (NExScI), under contract with the National Aeronautics and Space Administration (PI A.Boesgaard).
 }
}
\author {
Spite, M.\inst{1}\and
Bonifacio, P.\inst{1}\and 
Spite, F.\inst{1}\and
Caffau, E.\inst{1}
\and
Sbordone, L.\inst{2}
\and
A.\,J.\,Gallagher \inst{3,1}
}
       
\institute {
GEPI, Observatoire de Paris, Universit\'{e} PSL, CNRS,
Place Jules Janssen, 92190 Meudon, France
\and
European Southern Observatory, Casilla 19001, Santiago, Chile
\and
Max Planck Institut f\"ur Astronomie, D-69117 Heidelberg, Germany
}


\authorrunning{Spite  et al.}

\titlerunning{Abundance of Be in an UMP dwarf}

  \abstract
{Measurable amounts of Be could have been synthesised primordially if
the Universe were non-homogeneous or in the presence of late decaying
relic particles.}
{We investigate the Be abundance in the extremely metal-poor star \JJ
([Fe/H]=--3.84) with the aim of constraining inhomogeneities or the
presence of late decaying particles.     
}
{High resolution, high signal-to-noise ratio UV spectra were acquired
at ESO with the Kueyen 8.2 m telescope and the UVES spectrograph.
Abundances were derived using several model atmospheres and spectral synthesis code.
}
{ We measured  
$\rm \log(Be/H) = -14.3$ from a spectrum synthesis of the region of the Be line. 
Using a conservative approach, however we adopted an upper 
limit two times higher, i.e. $\rm \log(Be/H) < -14.0$. We measured the O 
abundance from UV OH lines and find [O/H]=--3.46 after a 3D correction. }
{Our observation reinforces the existing upper limit on primordial
Be.  There is no observational indication for a primordial production of 
$\rm ^9 Be$.  This
places strong constraints on the properties of putative relic
particles.  This result also supports the hypothesis of a homogeneous
Universe, at the time of nucleosynthesis.  
Surprisingly, our upper limit of the Be abundance is well below the
Be measurements in stars of similar [O/H]. This may be evidence that
the Be-O relation breaks down in the early Galaxy, perhaps due to
the escape of spallation products from the gas clouds in which
stars such as \JJ have formed.  
}

\keywords{ Stars: Abundances -- Galaxy: abundances --  Galaxy: halo -- 
stars: individual: 2MASS J18082002-5104378 }

\maketitle

%
\section{Introduction}

Beryllium has only one stable isotope, $^9$Be.  The lack of stable
nuclei with A=5 implies that it cannot be synthesised by capture of
$\alpha$ particles and the lack of stable nuclei with A=8 implies it
cannot be synthesised by proton nor neutron captures either.  The
nuclear fusion reactions that can synthesise Be in a plasma involve
several rare nuclei, specifically :
 $\rm ^7Li(^3H,n)^9Be$,
 $\rm ^7Be(^3H,^1H)^9Be $, and
$\rm ^6He(^4He,n)^9Be. $

In the conditions that characterise stellar interiors (including H and
He burning shells) such reactions cannot synthesise Be fast enough to counter the
inverse photodissociation reactions that destroy this element.  Thus stars
are net destroyers of Be.

In the first minutes of existence of the primordial plasma 
when most of the helium in the Universe was produced very tiny
amounts of $^9$Be can be formed; this can occur at the level of log($^9$Be/H)$\approx
10^{-18}$ \citep{Pitrou}, that is almost eight orders
of magnitude less than the primordial $^7$Li.  
This is however true under ``standard''
conditions, that is if the plasma is homogeneous and there is no ``new
physics''.  If the primordial plasma was inhomogeneous and in
particular included lower density $n$-rich regions,
\citet{BoydKajino} showed that $\rm^3H$ and $\rm^7Li$ could be
abundant enough to make the $\rm ^7Li(^3H,n)^9Be$ reaction, which is an
efficient channel to produce sizeable amounts of $^9$Be.  A way to
introduce new physics is to postulate the existence of relic
particles, interacting either electromagnetically or strongly, which
decay at late times \citep[see e.g.][and references
therein]{Jed06,Kusakabe}.  For example, \citet{Pospelov} showed
that the energy injected by decaying hadrons can lead to an efficient
$^9$Be production via the $\rm ^6He(^4He,n)^9Be $ reaction.  In fact
they advocate the use of an upper limit on the primordial $^9$Be
abundance as a powerful test to put limits on the energy and decay
half-life of such putative relic hadrons. The common wisdom, supported 
by the observations (see below) is that all the observed Be 
is produced by spallation processes triggered by
cosmic rays in the interstellar medium \citep{Reeves,Meneguzzi}.

From the observational point of view, Be can be observed in solar-type
stars via the \ion{Be}{ii} resonance doublet at 313\,nm.  This
makes the observation from the ground difficult since this wavelength
 is rather near to the atmospheric cut-off.  The Be abundance in the Sun
was determined using the lines from \citet{Chm75} who derived a Be
abundance about 0.3\, dex lower than the meteoritic abundance.
Coupled to the fact that soon after \citet{Boesgaard76} found almost
the same Be abundance in a sample of young stars, this uniformity led to the
notion that Be is depleted in the Sun, like Li.  The solar abundance
of Be was drastically revised by \citet{BB98} who invoked the presence
of an unaccounted 
continuum opacity at UV wavelengths and derived
a Be abundance that is in good agreement with the meteoritic
abundance.  The motivation for this extra opacity was to force the UV
and IR lines to yield the same O abundance.  An analysis of these
lines using 3D hydrodynamical simulations by \citet{Asplund04}
confirmed the need for this extra opacity.  It should be noted however 
that the source of this continuum opacity has not to date been identified 
and that it is not unanimously accepted \citep[see e.g.][]{BK02}.  
Recently \citet{Carlberg}, using a new line list in the near UV 
for generating theoretical solar spectra in the region of the Be lines, 
found that the difference in Be abundance is only 0.2\,dex 
with or without an extra opacity. This implies that 
even using this extra opacity 
Be is depleted by about 0.1\,dex in the solar photosphere.

The first attempts to measure Be in Pop II stars to study the
Galactic evolution of Be began in the 1980s \citep{MB84,Molaro},
however it was not until the late 1980s and 1990s that it became
clear that the Be abundance shows a clear linear decrease with
decreasing metallicity
\citep{RMAB88,Gil91,Ryan92,Gil92,BK93,Molaro97}.  If Be is produced
only by cosmic rays, then the Be abundance can be used as chronometer,
provided there is a suitable model of the temporal evolution of the
cosmic ray evolution \citep{Beers00,Pasquini05}.  The advent of 8\,m
class telescopes with high resolution spectrographs that can observe
down to the atmospheric cut-off allowed the measurement of Be in a large
sample of field halo stars
\citep{Boesgaard99,Primas00a,Primas00b,Boesgaard07,SmiljanicPB09,Ito,Tan09,Tan11,BoesgaardRL11}
and also in two globular clusters \citep{Pasquini04,Pasquini07}.

Be abundances in metal-poor stars allow for probing the existence of
inhomogeneities in the primordial Universe or the existence of late
decaying relic particles.  If there is no primordial production of Be, 
the linear decrease of the Be abundance with decreasing metallicity 
should continue no matter how low the metallicity of the star.  
If instead there is primordial production of Be, at some metallicity value 
the Be abundance should stop decreasing and present a constant value at all lower metallicities below.  
Thus measurements and upper limits of Be at the lowest abundances 
are of paramount importance to probe a primordial production of Be.  
The discovery of the bright extremely
metal-poor star \JJ ([Fe/H]=--3.8) by \citet{MelendezPT16} opens up
the possibility to probe the Be abundance in stars at the lowest
metallicities.  In this paper we present the analysis of high
signal-to-noise ratio (S/N) UV spectra of the star \JJj \  acquired with the
specific aim of investigating its Be abundance.

\section {Observational data} 

In order to observe the Be line at 313\,nm, new spectra of the ultra
metal-poor (UMP) dwarf \JJ were obtained in June 2018 with the Very Large Telescope (VLT) and the spectrograph UVES \citep{DekkerDK00}.  
Ten 1 h exposures were obtained during the night 
of June 21-22.
The dichroic beam-splitter was used permitting simultaneous use of the 
blue and red arms.  The blue arm was centred at 346\,nm and the red
arm at either 760 or 860\,nm.  With the spectra obtained previously
with UVES by \citet{MelendezPT16}, the spectral coverage of this UMP dwarf is almost complete from 310 nm to 1\,000 nm (with
only a gap between 452.3 nm and 478.6 nm).  The resolving power $R$ is
close to 50\,000\relax\ in the blue and 40\,000\relax\ in the red.  In
the region of the Be doublet (313\,nm), the S/N of the spectrum is close
to 70  a value close to the expected value in case of good weather 
(seeing of 1'' and good  transparency), it is about 250 at 370\,nm 
and 350 at 670\,nm. 

The spectra were reduced using the ESO UVES pipeline version 5.8.2;
the basic concepts and algorithms of the pipeline can be found in
\citet{BallesterMB00} and in the user manual.  The spectra were
extracted using optimal extraction and flat-fielding was performed on the
extracted spectra.  Two different flat-field lamps were used: a
deuterium lamp below 340\,nm and a tungsten lamp, for longer
wavelengths.  The spectra were wavelength calibrated using the Th-Ar
lamp exposures.

We carefully measured the radial velocity on our spectra and on the previous UVES spectra \citep{MelendezPT16}.  The more precise measurement of radial velocities with UVES is obtained when stellar and telluric lines are present in the spectrum.  The zero point of the wavelength scale depends indeed on the position of the star on the slit \citep[see e.g.][]{Molaro08} and the position of the telluric lines on the spectrum makes its definition possible.\\  
In very metal-poor stars
this is possible only on the yellow spectra domain centered at 580\,nm.
Unfortunately the determination of the zero point is not possible on
the spectra centered at 346\,nm since there are no telluric lines in
this region.  It is possible to determine the zero point on the
spectra centered at 760\,nm, but in this wavelength range, the \Feu~
lines in the stellar spectrum are extremely weak and only the position of the hydrogen line \Ha~ and of the red \Cad~ triplet could be measured.  
In the blue and in the visible region (settings B346 and R580 in table \ref{vrad}) the wavelength of the stellar iron lines were compared to the wavelength of numerous \Feu~ lines taken from the list of \citet{NaveJL94}. The wavelengths of the telluric lines are from \citet{Jacquinet-HussonSC05}.  
The velocity error on the barycentric radial velocity in Table \ref{vrad} should be less than 1.0\,\kms. The star \JJ has been observed by Gaia DR2 \citep{GaiaDR2} but its radial
velocity is not provided.

\section{Binary nature and orbit}

\citet{Schlaufman18} confirmed the binary nature of \JJj.
Using 17 radial velocity measurements from spectra obtained
with MIKE at the Magellan telescope, and three measurements
obtained from the UVES  R580 spectra (which we also used), \citet{Schlaufman18} were able to determine the orbital parameters for this system. They also gathered 31 epochs of radial velocity measurements obtained from low resolution spectra using GMOS-S on the Gemini South telescope. 
Since we have independent measurements of the radial velocities 
(the UVES  R580 spectra and a new epoch from our UVES R760 spectrum),
we decided to redetermine the orbital parameters of the system
combining our measurements with those of \citet{Schlaufman18}.

\begin{figure}
\begin{center}
\resizebox{7.5cm}{!}  
{\includegraphics[clip=true] {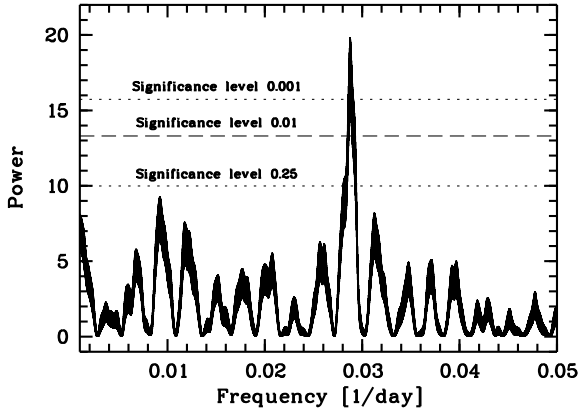}}
\caption[]{Lomb-Scargle estimate of the power spectrum of all the radial velocity measurements.
The lower dotted line corresponds to a false alarm probability of 0.25, dashed line to 0.01., and upper dotted line to 0.001.}
\label{power_all}
\end{center}
\end{figure}

\begin{figure}
\begin{center}
\resizebox{7.5cm}{!}  
{\includegraphics [clip=true]{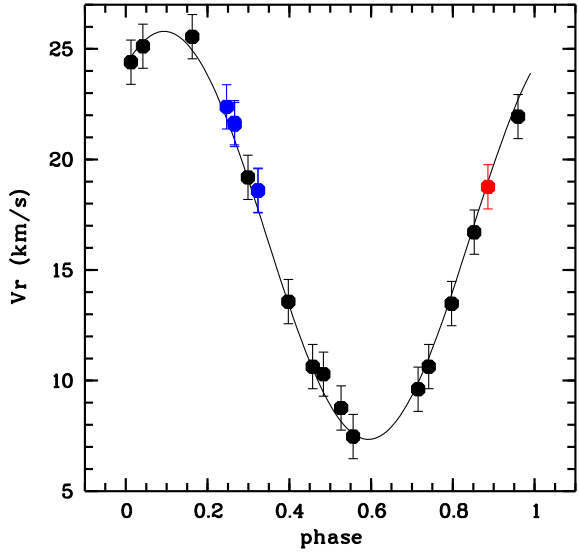}}
\caption[]{Computed orbit for the \JJ system (using
the parameters in Table \ref{orbit_param}) compared with the 
observed radial velocities based on high resolution spectra.
The black points indicate the measurements of \citep{Schlaufman18}.
The blue points indicate our measurements based on the UVES R580 spectra
given in Table\,\ref{vrad}. The red point
indicates our measurement based on the UVES R760 spectrum as given in Table\,\ref{vrad}.}
\label{plot_orbit}
\end{center}
\end{figure}

In our opinion the most robust determination of the
period of a binary system comes from the power spectrum
of the observations. 
To estimate the power spectrum of the radial
velocities measuremnts we used the Lomb-Scargle periodogram
\citep{Lomb,Scargle}.
If we use only the measurements based on high resolution spectra,
i.e. MIKE and UVES, no peak is statistically significant; in that case, all the peaks appearing
could be due to random noise.
In Fig.\,\ref{power_all} we show the power spectrum
obtained from all the radial velocity measurements, 
including those based on the GMOS-S spectra.
In this case, a highly significant peak, which has a false
alarm probability less than 0.001,  
is apparent corresponding to a period of 34.7538 days. 
This period is almost identical to that obtained
by \citet{Schlaufman18} using Keplerian fits to their 
high resolution data. 
We decided to fit a Keplerian orbit to our 
radial velocities measurements based on high resolution spectra, keeping
the value of the period fixed. 
We used version 1.3 of the program {\tt velocity}
\citep{Wichmann}.
Our preferred solution is summarised in Table\,\ref{orbit_param}
and is very close to that found by  \citet{Schlaufman18}
except for the angle of the periastron, the time of passage at periastron and the eccentricity. We did not run a Monte Carlo to estimate errors, since this orbit is certainly preliminary. The star is bright enough that it should eventually have  radial velocities for about 80 epochs from the RVS spectrograph on board Gaia \citep[see e.g.][]{Sartoretti}. 
The ensemble of ground-based and space-borne radial velocities
will provide a much more accurate orbit. The orbit and the phased
data for high resolution measurements are shown in Fig.\,\ref{plot_orbit}, where we assumed an error of 1\,\kms\
for all the measurements. The root-mean-square deviation of our computed
orbit from the observations is 0.52\,\kms.

\begin{table}
\begin{center}   
\caption[]{
Radial velocity of \JJj. }
\label{vrad}
\begin{tabular}{l@{~~}c@{~~}c@{~~}c@{~~}c@{~~}c@{~~}c@{~~}c@{~~}c@{~~}
c@{~~}c@{~~}c@{~~}c}
\hline
  Date      & MJD         &  RV    &  Bary. &   RV  &    RV   & sigma\\
            &             & (geo.) &  corr. &(tell.)& (bary.) &      \\
\hline
\multicolumn{2}{l}{{\bf R580 spectra} }\\
19-10-2014  & 56949.01289 & 46.50&  -24.22&  0.62&  21.66&  0.26& \\
19-10-2014  & 56949.02388 & 46.63&  -24.23&  0.82&  21.58&  0.18& \\
21-10-2014  & 56951.00724 & 42.79&  -23.84&  0.37&  18.58&  0.25& \\
21-10-2014  & 56951.01822 & 42.98&  -23.85&  0.52&  18.61&  0.11& \\
06-03-2015  & 57087.35943 & -3.38&  +25.63& -0.13&  22.38&  0.13& \\
06-03-2015  & 57087.37042 & -3.05&  +25.62& -0.03&  22.60&  0.21& \\
\\
\multicolumn{2}{l}{{\bf B346 mean spectrum} }\\
21-06-2018  & 58291.03014 & 17.71&   +0.72&  -   &  18.43&  0.12&\\
\\
\multicolumn{2}{l}{{\bf R760 mean spectrum} }\\
21-06-2018  & 58291.03014 & 18.28&   +0.72&  0.25&  18.75&   -- &\\ 
\hline    
\end{tabular}  
\end{center}   
\end{table}

\begin{table}
\begin{center}   
\caption[]{
Orbit parameters for the system \JJj. }
\label{orbit_param}
\begin{tabular}{lr}
\hline
Radial velocity of the barycentre & 16.57 \kms \\
Radial velocity amplitude         &  9.23 \kms \\
Eccentricity                      &  0.00 \\
Angle of the periastron           &  -33.71 degrees \\
Period                            & 34.7538  days \\
Time of passage at periastron     & 2456453.72 HJD \\
\hline    
\end{tabular}  
\end{center}   
\end{table}

\section {Stellar parameters} 
\citet{MelendezPT16} estimated the
temperature of \JJ by imposing the excitation equilibrium of \Feu~
lines and adding an empirical correction described in
\citet{FrebelCJ13}.  Since \JJ was observed by Gaia, we used the Gaia
photometry recently displayed in the Gaia DR2
\citep{ArenouLB18,GaiaDR2}, and the 3D maps of interstellar reddening
\citep{CapitanioLV17,LallementCR18,Lallementpriv} to improve these
parameters.  The Gaia photometry and the reddening are listed in
Table \ref{gaiadata}.
We note that following \citet{Schlaufman18}, the mass of the secondary must be very low ($M_{2}=0.14 M_{\odot}$) and thus its contribution to the total flux is negligible.

\begin{table}
\begin{center}   
\caption[]{
Photometry and distance of \JJj ~~(Gaia DR2 6702907209758894848). The parallax of the star, parallax error, observed $g$ magnitude, and $(BP-RP)$ colour are given in the first row; next the 
distance of the star in pc with the minimum and maximum values of this distance taking into account the parallax error is given.
The extinction in the G magnitude and 
$E_{(BP-RP)}$ are found in the second row. Then the values $g_{0}$ and $(BP-RP)_{0}$ are the values of $g$ and $(BP-RP)$ corrected for the reddening, and G is the magnitude corrected  for extinction and distance (absolute magnitude). 
 }
\label{gaiadata}
\begin{tabular}{l@{~~}r@{~~}r@{~~}c@{~~}c@{~~}l@{~~}c@{~~}c@{~~}c@{~~}
c@{~~}c@{~~}c@{~~}c}
\hline
Parallax&  Par. & Obs.    &$(BP-RP)$&   dist.& dist. & dist.\\
 ms     & error & $g$ mag.& mag.   &    (pc)& min   & max  \\
\hline                                                                                     
 1.6775  & 0.0397& 11.756 & 0.903 &  596   & 582   & 611   \\
\hline
\\
 A(G) &$E_{(BP-RP)}$& $g_{0}$ &$(BP-RP)_{0}$ & G abs  \\
=A(V) &  mag.   & mag.    & mag.         &  mag.  \\
\hline                    
 0.210& 0.105   &  11.546 &  0.798  & $2.67 \pm 0.05$   \\  
\hline
\end{tabular}  
\end{center}   
\end{table}

\begin{figure}
\begin{center}
\resizebox{6.0cm}{7.0cm}  
{\includegraphics {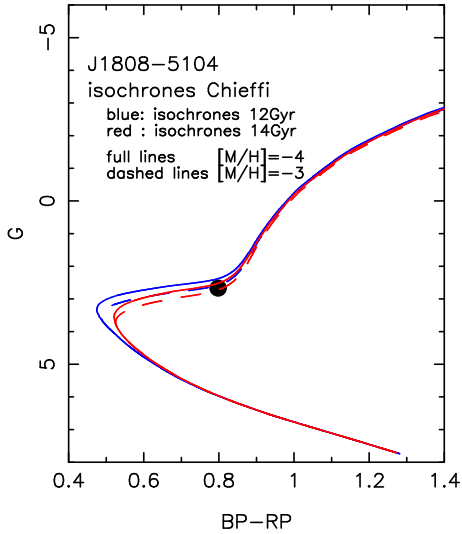}}
\caption[]{Position of \JJ (black dot) in a G vs. (BP-RP) diagram and
comparison to isochrones computed by Chieffi \& Limongi for very
metal-poor stars.}
\label{iso}
\end{center}
\end{figure}

In Fig.\,\ref{iso} we compared the position of \JJ in a G versus
(BP-RP) diagram to the isochrones computed by Chieffi\& Limongi
\citep{Chieffipriv}; we used  the same code and prescriptions as
\citet{SCL97} for 12 and 14 Gyr and metallicities of $-3.0$ and
$-4.0$.  We note that the error on the G magnitude is very small and is
inside the black dot in Fig.\,\ref{iso}.  The position of the star in the diagram
corresponds to a subgiant star with \Teff=5\,600\,K, log\,g=3.4, 
$ M\approx0.8M_{\odot}$, and we decided to adopt this model.  
 As a check we also computed the \Ha~ profile for the 1D model adopted by 
Mel\'endez (\Teff=5\,440\,K, log\,g=3.0) and the model adopted in this work (\Teff=5\,600\,K, log\,g=3.4). The fit of the wings of \Ha~ is better with our model. The use of 3D profiles would even point towards a slightly hotter temperature \citep{AmarsiNB18}.

\section{Analysis}
\begin{figure}
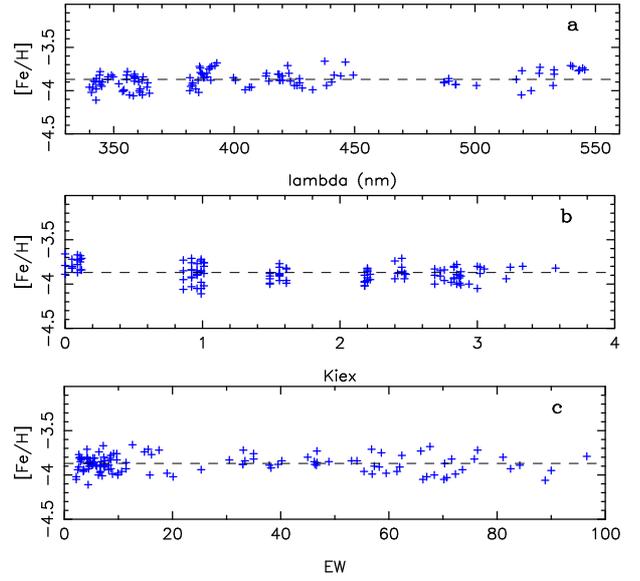

\begin{center}
\resizebox{8.0cm}{2.5cm}  
{\includegraphics {kivplotplezJ1808-5104-a.ps}}
\resizebox{8.0cm}{2.5cm}  
{\includegraphics {kivplotplezJ1808-5104-b.ps}}
\resizebox{8.0cm}{2.5cm}  
{\includegraphics {kivplotplezJ1808-5104-c.ps}}
\caption[]{Iron abundance from individual lines in \JJ as a function
of the wavelength, excitation potential, and, equivalent width
of the line.}
\label{kiv}
\end{center}
\end{figure}

Since the adopted atmospheric model is rather different from the model adopted by \citet{MelendezPT16}, we redetermined the abundances of the different elements with our adopted model.  We carried out a classical
Local Thermodynamical Equilibrium Analysis (LTE) analysis using OSMARCS models
\citep{GustafssonBE75,GustafssonEE03,GustafssonEE08}. 
The abundances
were derived using equivalent widths or fits of synthetic spectra
when the lines were blended.  We used the code {\tt turbospectrum}
\citep{AlvarezP98}, which includes treatment of
scattering in the blue and UV domains.  These abundances are given in
Table \ref{abund}.  In Fig.\,\ref{kiv}, we show the dependence of the iron abundance on the wavelength, excitation potential, and equivalent width of the \Feu~ line.

\noindent--In Fig.\,\ref{kiv}\,b the iron lines with an excitation potential close to zero, over-predict the iron abundance. This is because of non-LTE (NLTE) effects, so these lines are not taken into account in the determination of the mean iron abundance \citep[see Fig.\,3 in][]{CayrelDS04}.\\
--In Fig.\,\ref{kiv}\,c  the iron abundance does not depend on the equivalent width of the line, and thus it justifies the choice of the microturbulent velocity, i.e. \vt\,=\,1.6\,\kms.\\
The final 1D, LTE abundances are given in Table \ref{abund}.  The
solar abundances are from \citet{CaffauLS11} or \citet{LoddersPG09}.


\begin{table}
\begin{center}   
\caption[]{
Derived abundances in \JJ for \Teff=5600\,K, log g=3.4, \vt=1.6\,\kms (1D computations, OSMARCS model). }
\label{abund}
\begin{tabular}{l@{~}r@{~}r@{~}r@{~~~~}r@{~~}r@{}c@{~~}c@{~}c@{~~}
c@{~~}c@{~~}c@{~~}c}
\hline
 Elem.  &$A(X)_{\odot}$&$A(X)_{\star}$& N& $\sigma$& [X/H]& ~[X/Fe]]  \\
\hline
  \Feu   &  7.52   &  3.63    & 65   & 0.08  &               &             \\
  \Fed   &  7.52   &  3.73    &  4   & 0.07  &               &             \\
Fe adopt.&  7.52   &  3.68    &      &       &    --3.84 &             \\
\hline  
  \Liu   &             &  1.78    &                                        \\
 C (CH)  &  8.50   &  5.15   &syn.   &       &    --3.35 & ~+0.49     \\
 O (OH)  &  8.76   &  6.28    &syn.   &      &    --2.48 & ~+1.36     \\
  \Nau   &  6.30   &  2.56    &  2   & 0.01  &    --3.74 & ~+0.10     \\
  \Mgu   &  7.54   &  4.16    &  6   & 0.06  &    --3.38 & ~+0.46     \\
  \Alu   &  6.47   &  1.93    &  2   & 0.04  &    --4.54 &~--0.70     \\
  \Siu   &  7.52   &  3.69    &  1   & 0.04  &    --3.83 & ~+0.01     \\
  \Cau   &  6.33   &  2.81    &  9   & 0.17  &    --3.52 & ~+0.32     \\
  \Scd   &  3.10   &--0.53    &  3   & 0.12  &    --3.63 & ~+0.21     \\
  \Tiu   &  4.90   &  1.67    &  8   & 0.05  &    --3.23 & ~+0.61     \\
  \Tid   &  4.90   &  1.47    & 15   & 0.13  &    --3.43 & ~+0.41     \\
  \Cru   &  5.64   &  1.57    &  5   & 0.12  &    --4.07 &~--0.23     \\   
  \Cou   &  4.92   &  1.80    &  4   & 0.10  &    --3.12 & ~+0.72     \\
  \Niu   &  6.23   &  2.59    &  3   & 0.04  &    --3.64 & ~+0.20     \\
  \Srd   &  2.92   &--1.86    &  2   & 0.04  &    --4.78 &~--0.94     \\
\hline
\end{tabular}  
\end{center}   
\end{table}

\subsection{Carbon and oxygen abundances} \label{CO3D}
In \JJj, the carbon abundance deduced from the CH band,  [C/Fe]=+0.49, is very close to the mean value found from 1D calculations for the extremely metal-poor turn-off stars \citep{BonifacioSC09}: [C/Fe]=+0.45. But the CH band is sensitive to 3D effects \citep{GallagherCB16}. We made use of a 3D \cobold model \citep{FreytagSL12} belonging to the CIFIST grid \citep{LudwigCS09}, with parameters (5\,500\,K/3.5/--4.0) close to the stellar parameters of \JJ to compute the 3D correction and we found ${\rm A_{\rm 3D}(C)}-{\rm A_{\rm 1D}(C)}=-0.40\pm 0.1$\,dex; the error in this case corresponds to the different estimations of this correction in different parts of the CH band.

The oxygen abundance is derived from a fit of the ultraviolet OH band between 312.2 and 313.2\,nm. The uncertainty (scatter from line to line) is less than 0.1 dex.
The OH band is also strongly affected by 3D effects.   For
12 OH lines, we computed the 3D corrections \citep{zolfito} and we
derived ${\rm A_{\rm 3D}(O)}-{\rm A_{\rm 1D}(O)}=-0.98\pm 0.08$\,dex.
As a consequence in \JJj, A(C)=4.75, [C/H]=--3.75,
[C/Fe]=+0.09 and  A(O)=5.30 with [O/H]=--3.46
and [O/Fe]=+0.38.

\subsection{Na and Al abundances}
The abundance of these two elements in Table \ref{abund} has been
deduced from the resonance lines which are often affected by strong
NLTE effects.  We estimated the NLTE correction from
\citet{AndrievskySK07,AndrievskySK08}.  
In fact the correction at this very low metallicity is small for \Nau.  We found $\rm\Delta(A(Na))$ =--0.05, but the correction is large for \Alu\,, i.e. $\rm\Delta(A(Al))$ =+0.68.  In \JJj, the abundances of Na and Al given in Table \ref{abund}, after correction for NLTE effects, are A(Na)= 2.51, [Na/Fe]=+0.05 and A(Al)=2.61 and [Al/Fe]=--0.02

\begin{figure}
\begin{center}
\resizebox{\hsize}{!}                   
{\includegraphics {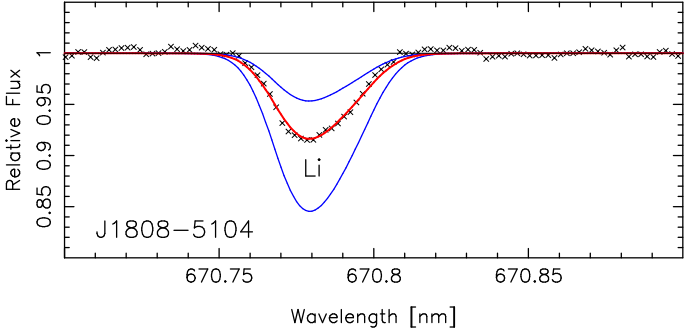}}
\caption[]{Observed spectrum of \JJ in the region of the Li doublet and 
synthetic spectra computed with A(Li)=1.5 and 2.1 (blue lines) and 1.78 (red 
line, best fit). 
}
\label{lifig}
\end{center}
\end{figure}

\begin{figure}
\begin{center}
\resizebox{\hsize}{!}                   
{\includegraphics {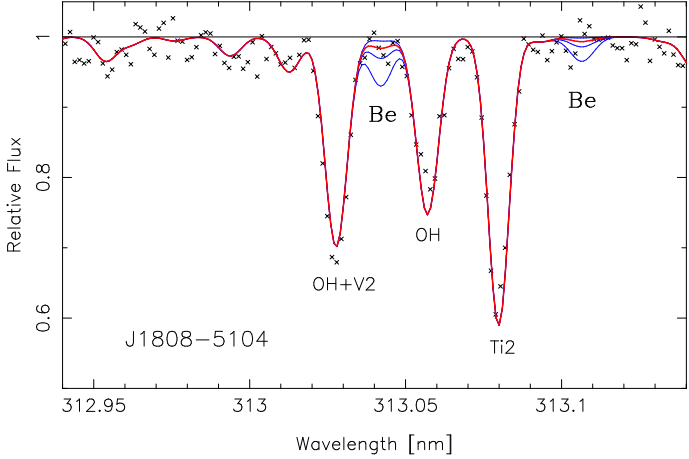}}
\caption[]{Observed spectrum of \JJ in the region of the Be doublet and 
synthetic spectra computed with A(Be)=--3.0, --2.0 and --1.6 (blue lines), and 
--2.33 (red line, best fit). Taking into account the uncertainty in the position 
of the continuum, we adopted $\rm A(Be)\leq -2.0$ dex i. e. $\rm \log(Be/H) \leq -14.0$.
}
\label{befig}
\end{center}
\end{figure}

\section{Li and Be abundances}
The Li abundance is redetermined from our new high S/N spectra in 
the red (Fig. \ref{lifig}).
We find A(Li)=1.78 a value very close to the Li abundance found by 
\citet{MelendezPT16}. 
If we apply the 3D-NLTE correction computed by \citet{Sbordone10} we find A(Li)=1.88. 
As a consequence Li is slightly depleted  below the 
Spite plateau value \citep{Spite82a,Spite82b,Sbordone10}. 
Lithium indeed is a very fragile element; it is destroyed by proton fusion when temperature reaches about $2~10^{6}$\,K. In a main sequence star this element is destroyed as soon as the convective zone reaches the layers where the temperature is higher than  
this fusion temperature. 
When the star leaves the main sequence it develops surface convection zones, which deepen as the star evolves to lower temperatures. The surface convection zone mixes the surface layer with deeper material in which lithium has been depleted, and the observed lithium abundance falls. Following \citet{PilachowskiSB93}, the decrease of  Li abundance for \Teff = 5600\,K is  $-0.25\,\pm 0.25$ dex in excellent agreement with the observed Li abundance in \JJj. 

 The abundance of Be is determined by a  $\chi^{2}$ fit of the observed spectrum (Fig.\,\ref{befig}) between 312.94 and 313.14\,nm. The best fit depends on the adopted position of the continuum, which is fortunately rather well defined in this region. Only the bluest Be line is used, as is generally done in metal-poor stars, the reddest being much too weak. The Be line at 313.04\,nm is close to two OH lines, and to compute the profile of the global feature we use the 1D oxygen abundance given in Table \ref{abund}.  
The best fit is obtained for A(Be)=--2.33, but considering the S/N of the spectrum, it is reasonable to say that  A(Be)< --2.0 or log(Be/H)<--14.0. 
Since the weak \Bed~ lines are formed in the deep atmospheric layers,  the abundance of Be computed with the 1D-LTE or the 3D-NLTE hypotheses are not significantly different \citep{Primas00b}.

Although Be is destroyed at a higher temperature than Li ($3.5~10^{6}$\,K), it is legitimate to ask whether Be has been depleted in \JJj, since its Li abundance is slightly below the ``plateau''.  
However, we expect that for a same phase of the evolution of the star and thus the same depth of the convective layer, the depletion of Be by dilution is much smaller than the depletion of Li. In the sample of \citet{BoesgaardRL11} there is no significant difference between the ratios [Be/Fe] in turn-off and in turn-off and insubgiant stars.

\section{Discussion}

\subsection{Be-Fe correlation}

In Fig.\,\ref{befe-trend} we plot log(Be/H) versus [Fe/H] for a large sample of Galactic stars from \citet{SmiljanicPB09} and
\citet{BoesgaardRL11}.  When a star was in both lists we prefer the abundance given by \citet{SmiljanicPB09}, but the two are
always very close.
The black dashed line in Fig \ref{befe-trend} represents the regression 
line in the middle of these stars.

The two  Be measurements at the lowest metallicity are for the stars G 64-12 and G 275-4

\begin{figure}
\begin{center}
\resizebox{\hsize}{!} 
{\includegraphics{abfe-be-smil-boes2-J1808.ps}} 
\caption[]{For galactic stars, log(Be/H) vs. [Fe/H].  The green open circles are from
\citet{SmiljanicPB09} and the blue filled circles from
\citet{BoesgaardRL11}. However the position of G\,64-12 in this diagram takes into account our new measurement of the Be abundance in this star adopting the model of \citet{Primas00b}, which  is in better agreement with the Gaia-DR2 data. 
The upper limit of the abundance of Be in \JJ and \BD~ are indicated with 
big red and blue open circles.  The blue dashed straight line represents the mean relation.  The curved red dash-dotted line at low metallicity represents the possibility of a plateau, suggested in particular, by the previous  high Be abundance found in G\,64-12 by \citet{Primas00b} and \citet{BoesgaardRL11}. The very low Be abundance in the two additional stars \BD~ and \JJ rules out the possibility of a plateau.}
\label{befe-trend}
\end{center}
\end{figure}

\begin{figure}
\begin{center}
\resizebox{\hsize}{!}                   
{\includegraphics {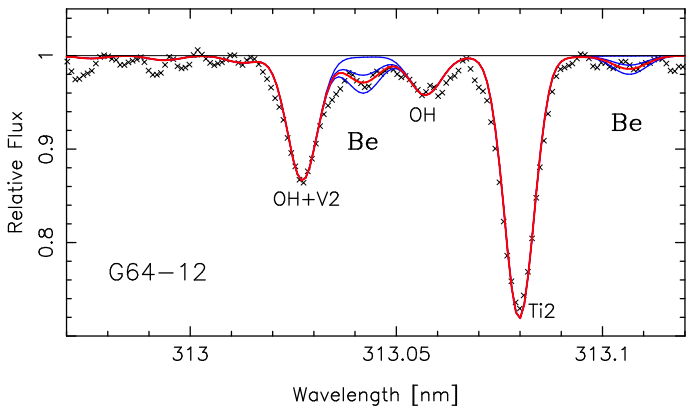}}
\caption[]{Mean of the reduced G64-12 spectra obtained with the HIRES-Keck and the UVES-VLT spectrographs in the region of the Be doublet and 
synthetic spectra computed with A(Be)=--3.0 --1.5 --1.2 (blue lines), and 
--1.35 (red line, best fit). } 
\label{BeG6412}
\end{center}
\end{figure}

The two Be measurements at the lowest metallicity are for the stars
G\,64-12 \citep{Primas00b,BoesgaardRL11} and G\,275-4
\citep{BoesgaardRL11}. 
For G\,64-12 \citet{Primas00b}
measure $\rm log(Be/H)=-13.10\pm 0.15,$ while \citet{BoesgaardRL11}
found $-13.43\pm 0.12$.  The two values are compatible within errors,
however the difference can be entirely explained by the different
atmospheric parameters adopted.  
\citet{Primas00b} indeed adopted \Teff = 6\,400\,K, \logg = 4.1, while \citet{BoesgaardRL11} adopted \Teff = 6074, \logg = 3.75. This relatively high Be abundance in G\,64-12 suggested at that time the possible existence of a ``plateau'' of the Be abundance at very low metallicity.

As a check we retrieved the UVES and HIRES spectra from the archives. In the region of the Be line, they have almost the same resolution and same S/N so it is possible to average these spectra. The resulting spectrum has a S/N of more than 150 in the region of the Be lines.
On this averaged spectrum we remeasured the O and Be abundance in  G\,64-12 adopting the Primas's model. 
The Be abundance is indeed very sensitive to surface gravity and the Gaia DR2 parallax of G\,64-12 ($3.7626\pm 0.0856$)\,mas implies \logg = 4.25; this value is very close to the gravity adopted by \citet{Primas00b}.
We found that the best fit was obtained with A(Be)=--1.35 or $\rm log(Be/H)=-13.35$ (Fig. \ref{BeG6412}). This value is intermediate between the \citet{Primas00b} and \citet{BoesgaardRL11} values. In Fig. \ref{befe-trend} the dot representing G\,64-12 takes into account this new measurement. The position of the very metal-poor stars  G\,64-12, G\,64-37, and G\,275-4 in this figure could suggest the existence of a plateau at a level of  $\rm log(Be/H)\approx-13.4$.

But before the present work, the most stringent upper limit on the primordial Be abundance was already the upper limit found in the carbon enhanced metal-poor  (CEMP) star BD\,$+44^\circ 493$ 
([Fe/H]=--3.7) provided by \citet{Ito}, i.e. log(Be/H)$< -14.0$.  The upper 
limit that we have derived for \JJ is essentially equivalent to 
that provided by BD\,$+44^\circ 493$.  The two stars have similar
atmospheric parameters and metallicities, however there is a big
difference in their carbon and oxygen abundances.  With reference to the classification introduced by \citet{toposIV}, while BD\,$+44^\circ493$ is a low-carbon band CEMP star, \JJ
is a carbon normal star.  This finding confirms what was already noticed
by \citet{Ito}: there is no correlation between CEMP nature and Be abundance.  

The very low abundances of Be in \JJ 
confirms that the possibility of a Be plateau at a
level of $\rm log(Be/H)\approx-13.6$ is ruled out (Fig \ref{befe-trend}).  It seems
reasonable to assume that the Be abundance continues its linear
decrease with metallicity in the range --3.0 to   --4.0\,dex.
We found that, on average, log $\rm(Be/H) = -10.75 + 0.89*[Fe/H]$ (see dashed line in Fig. \ref{befe-trend}).
The slope of this line is very close to the slope found by 
\citet{BoesgaardRL11} from their stars alone.
The low Be abundance found in \JJ and BD\,$+44^\circ 493$ is  
compatible with this linear regression.

The upper limit on the primordial Be abundance $\rm \log (Be_p/H) <
-14$ is thus reinforced as are the limits on late decaying hadrons
provided by \citet{Pospelov}.  At present there appears to be no hint
towards inhomogeneities in the primordial plasma or the presence of
late decaying particles.

\begin{figure*}
\begin{center}
\resizebox{8.0cm}{!} 
{\includegraphics{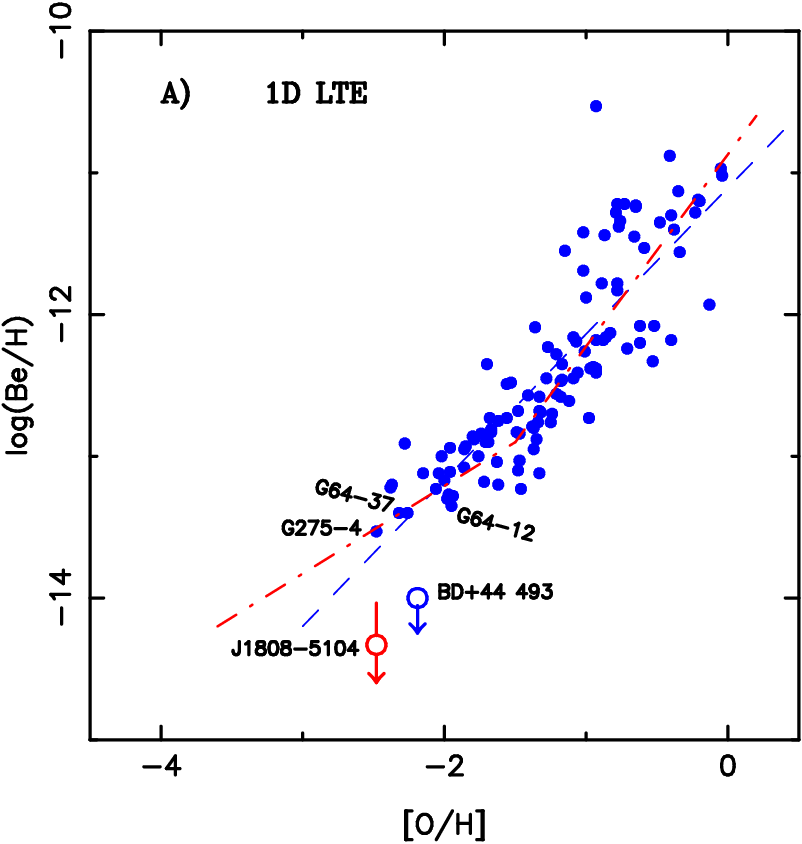}}
\resizebox{8.0cm}{!} 
{\includegraphics{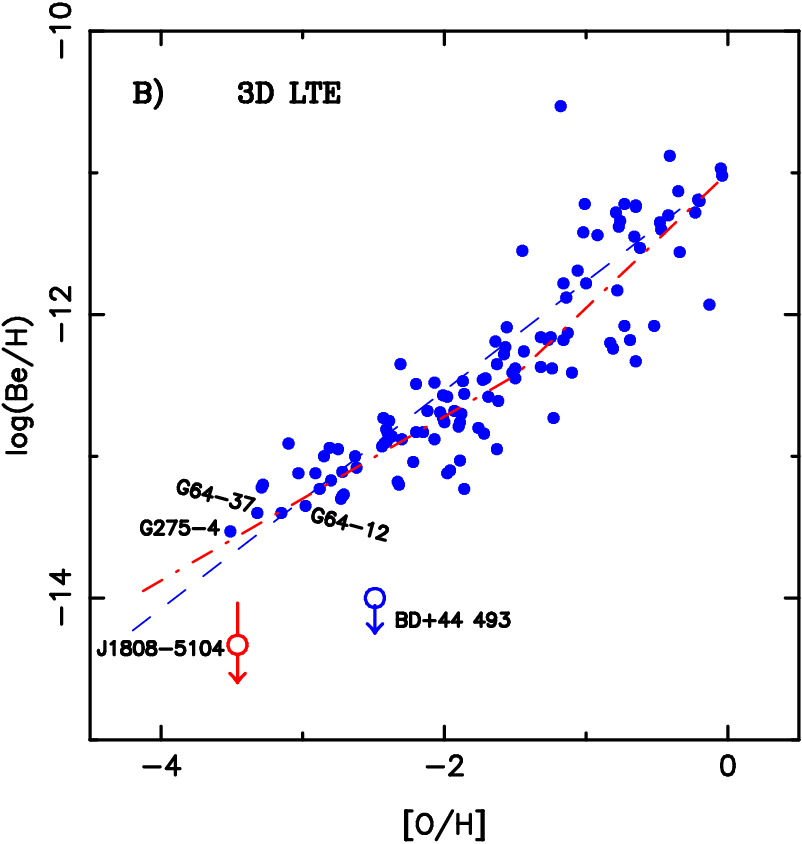}}
 \caption[]{For galactic stars, log(Be/H) vs. [O/H] following \citet{BoesgaardRL11} (blue filled circles).  The upper limit of the abundance of Be in \JJ and \BD~ are indicated with big red and blue open circles. In A) the oxygen abundance was simply computed with the 1D LTE hypothesis, as in \citet{BoesgaardRL11}, and in B) this abundance has been corrected for 3D effects. The blue dashed line represents the mean relation and the red dashed line the two-slope solution with a change of the slope around [O/H]=--1.5\,dex as in \citet{BoesgaardRL11}. 
The positions of \BD~ and \JJ in this diagram are hardly compatible with the mean relations.}
\label{beo-trend}
\end{center}
\end{figure*}

\subsection{Be-O correlation}

In Fig. \ref{beo-trend} we plott log(Be/H) as a function of [O/H] for the stars studied by \citet{BoesgaardRL11} in which the oxygen abundance was deduced from 1D computations of the profile of some UV-OH lines;  the authors estimated that the error in [O/H] is about 0.2 dex. As discussed in section \ref{CO3D}, these lines are strongly affected by 3D effects. 
The 3D correction has been computed by \citet{AsplundGP01} and \citet{GonzalezBL10} with two different grids of 3D models and their results are in good agreement. 
For dwarfs and subgiant stars this correction depends mainly on the metallicity and temperature of the star;  this correction
is negligible for $\rm[Fe/H]>-1.0$ but reaches almost --1.0 for $\rm[Fe/H]=-4.0$.  

The star \BD~ is peculiar, it is very C-rich and \citet{GallagherCB17} have shown that the carbon abundance affects the molecular equilibria in 3D hydrodynamical models in a much more prominent way than happens in 1D models. Using a C-rich hydrodynamical model atmosphere computed with the \cobold\ code 
\citep{FreytagSL12},
 we determined that the 3D correction in this case 
is only \mbox{--0.3 dex}. As a consequence [O/H] in \BD~ is equal to --2.49. 

The OH lines can be also affected by NLTE effects.
\citet{AsplundGP01}  approximated the UV-OH line formation with a two-level approach with complete redistribution, but they neglected the influence of NLTE on the dissociation of the molecules.
These authors found from 1D computations that the oxygen abundance derived from the UV-OH band is underestimated by about 0.2 dex, almost independently from the stellar parameters.
In order to obtain a right estimation of the NLTE effect it would be  necessary to take into account fully the 3D thermal structure of the  model atmosphere; this work is beyond the scope of this paper. 

To date, we neglected the NLTE correction and we applied to all the stars of the sample of \citet{BoesgaardRL11} the  3D correction computed from  \citet{GonzalezBL10}. 
\footnote {The 3D values of the oxygen abundance are given in a table available at the CDS.}

In Fig. \ref{beo-trend}A the oxygen abundance was computed from 1D models (as in \citet{BoesgaardRL11}) and in 
Fig. \ref{beo-trend}B,  this oxygen abundance was corrected for 3D effects.
If we had applied the \citet{AsplundGP01} NLTE correction, all the dots in Fig. \ref{beo-trend}B would have been shifted by 0.2\,dex towards higher [O/H] values.

When only Boesgaard's stars are considered, the data after 3D correction can be 
interpreted by a linear relation between log(Be/H) and [O/H] with a slope of 0.77. 
Alternatively a two-slope relation \citep[see ][]{BoesgaardRL11} with a slope of 0.95 \relax in the interval --1.5<[O/H]<0.0 and a slope of 0.58 at lower metallicity can be used.

In Fig. \ref{beo-trend}A or B, \JJ and \BD~ do not fit the general trend. 
The star \BD~ has an oxygen abundance close to the oxygen abundance of G\,64-12, and \JJ has about the same as  G\,275-4 and G\,64-37, but \JJ like \BD~ are clearly more deficient in Be.
The Be abundance expected in \JJj, from the mean relations in Fig.\,\ref{beo-trend}, would be in all the cases $\rm A(Be)\approx -1.6$, a value excluded from the observed spectrum (see Fig.\,\ref{befig}).
 
Since \BD~is a C-rich star it could be possible that the high CNO abundances in this star are the result of a  mass transfer from a ``now dead'' AGB companion. But since the star is a CEMP-no (no enrichment of the neutron capture elements) this interpretation is unlikely. 
Moreover following \citet{GaiaDR2}, the radial velocity of \BD~ does not seem variable, i.e.  RV=--147.9. 
As a consequence, the existence of a former pollution of the atmosphere of \BD~ by a massive companion in its AGB phase is questionable. It is highly probable that the abundance of C,N and O in the atmosphere of \BD~ is a good witness of the abundances in the cloud which formed the star.

\section{Conclusions}
 A two-slope solution of the relation log(Be/H) versus [O/H] is predicted by theory. It is now commonly accepted that all the observed Be is formed by spallation \citep{Reeves,Meneguzzi}, however the following two distinct processes can be invoked:
\begin{itemize}
\item H and He nuclei in cosmic rays  hit CNO nuclei in the ambient interstellar gas (secondary process).
\item CNO nuclei in the cosmic rays hit  H and He nuclei in the ambient interstellar gas (primary process).
\end{itemize}
If Be (and B) were formed preferentially by the secondary
process we would expect a quadratic dependence of the Be (B) abundance
on the oxygen abundance, hence a slope of two\relax\ in the logarithmic plane.
The primary process, on the other hand, would imply a slope of one, as
implied by the observations of both Be and B.
The secondary process  was probably invoked for the first time
as the main process for the production of B (and Be by extension)
by \citet{DuncanLL92} to explain their B observations. 
Further considerations on this point can be found in \citet{Duncan97,Molaro97}
and \citet{RichBoes09}. 
However, from the theoretical point of view \citet{Suzuki99} and \citet{Suzuki01}
argued that the primary process is the main source 
of all the observed B and Be. 

\citet{BoesgaardRL11} suggested that the balance shifted from primary to secondary in the course of time.
In the early days of Galactic evolution, the acceleration of 
CNO atoms from SNe II should be the main phenomenon and the number of 
Be atoms should be proportional to the number of SNe II and thus to the number of O atoms.
Later the number of O atoms is proportional to the cumulative number of SN II, 
while the energetic protons are proportional to the instantaneous number of SN II.
As a consequence the slope of the relation log(Be/H) versus [O/H] is expected to change, and for this reason \citet{BoesgaardRL11} tried to describe the Galactic evolution of Be with two straight lines and a break at a metallicity around --1.5. 
The slopes of the different relations in Fig. \ref{beo-trend}B are slightly different from those of \citet{BoesgaardRL11} since we correct the abundance of oxygen for 3D effects.

In \citet{toposII} we argued that low-carbon band CEMP stars, such as \BD,
were formed from gas that was polluted by SNe that experienced
a large fall-back of material onto the compact remnant, resulting in very high
ratios of CNO elements to iron. We referred to these  as ``faint supernovae'' (SNe) because we made the implicit assumption that the luminosities of these stars would also be lower than those of SNe that do not experience fall-back.
Observationally this same name is given to type II SNe that
are under-luminous, such as SN 1997D \citep{SN1997D}, which
was also characterised by relatively low expansion 
velocities and a low mass of ejected $\rm ^{56}Ni$ \citep{Turatto98}.
It is interesting to note that SN 1997D should not have been able
to produce light elements via spallation. 
The cross section for production of $\rm ^9Be$ via spallation
of oxygen drops drastically below energies of a few (MeV/A) 
of the projectile \citep[see figure 1 of ][]{Suzuki01}, and translated
into velocity of the O nuclei this requires velocities in excess
of 4000\,\kms. A typical type II SNe shows velocities of the ejecta
that are of the order of 10\,000\,\kms , thus in the useful range
for Be production. On the other hand SN 1997D showed an
expansion velocity of the ejecta of only 1\,200\,kms \citep{Turatto98},
which is clearly insufficient for Be production.
The upper limit of \BD\ seems to be consistent with the hypothesis
that it was formed from a faint SNe, characterised by strong fall-back,
responsible for the high CNO to Fe ratios, and low velocity
of the ejecta, resulting in a Be content that is clearly lower
than that of stars of similar [O/H].
In fact it could well be that \BD\ is completely devoid of Be
and that the fact that its upper limit on its Be abundance
falls exactly on the expected line of Be-Fe evolution is fortuitous.
It would be of paramount importance to be able to
push down the upper limit on the abundance of Be in \BD ,
even by only 0.3\,dex. 
As a corollary, measurements of Be in other lower carbon CEMP unevolved stars
are strongly encouraged. If our scenario is correct
all lower carbon CEMP unevolved stars should show a lower Be abundance
than carbon-normal stars of similar [O/H].
The unevolved lower carbon band CEMP star, HE\,1327-2326,
has an upper limit on the Be abundance log (Be/H)$<-13.2$
and a 3D corrected [O/H]=--2.64 \citep[][for this star the 3D correction is --0.72\,dex]{Frebel08}.   
This upper limit is inconclusive since
it is above the Be abundance measured in 
stars of comparable O abundance, and
also above both the two-slope model and the one-slope model.
It would be extremely important to be able to push
down this upper limit by 0.3 -- 0.4\,dex.
A detection of Be at the same level as that in
stars of similar O abundance
would invalidate our scenario on faint SNe.

While the low Be abundance in \BD~ could be interpreted as a result
of it being formed from ejecta of a faint SNe, the same
cannot be invoked for \JJj.
As we have already argued, the Li abundance measured in \JJ 
is strong evidence that the Be in this star cannot have been
significantly depleted.   
The fact that there are stars that have similar oxygen abundances
but significantly different Be abundances is a real puzzle. 
A fundamental piece of information would, of course, come
from a  measure of the Be abundance in \JJj or at least an 
upper limit lower by 0.3\,dex. This would at least
rule out the possibility that the difference is simply
due to large observational errors in both Be and O abundances. 
A far more intriguing possibility is that 
in the early Galaxy, the tight correlation of Be with O
breaks down: the scatter of the relation becomes larger.
In a simple picture, \JJ and \BD~ should be very old stars born in regions with 
anomalously high O, due to local inhomogeneities in the very early Galaxy. 

Given the very low metallicity of \JJ we may assume that 
the SNe, that have produced the metals that we observe in its
atmosphere, were Pop III stars, possibly even a single Pop III star 
or a few at most. If the velocities of the ejecta in these
stars were higher than observed in normal Pop I and Pop II  SNe, 
it may be that the lower mass spallation products, Li, Be and B,
escape with a high enough velocity to escape the cloud that 
will give rise to the next generation of stars, such as \JJj. 
If this were the case, we expect that stars at the metallicity
of \JJ ($\rm [Fe/H] \leq -3.5$) are all devoid of Be and B.
  
This would also be an interesting diagnostic to distinguish
true descendants of Pop III stars from stars of similar
metallicity, but formed from clouds polluted by
Pop II stars.
This clearly prompts for new observations: on the one hand,
it is important to measure Be and B in \JJ and other stars
of similarly low metallicity and, on the other hand,
one should increase the number of stars with measured Be and B
in the metallicity range [Fe/H]$\le -2.0$. A single or a few
Pop III SNe can pollute a gas cloud to such high metallicities and 
a measure of Be could allow us to detect such true Pop III descendants.

\begin {acknowledgements} 
AJG would like to acknowledge support by Sonderforschungsbereich 
SFB 881 ``The Milky Way System'' (subproject A5) of the German Research Foundation (DFG).
This work uses results from the European Space Agency (ESA) space
mission Gaia.  Gaia data are being processed by the Gaia Data
Processing and Analysis Consortium (DPAC).  Funding for the DPAC is
provided by national institutions, in particular the institutions
participating in the Gaia MultiLateral Agreement (MLA).  The Gaia
mission website is https://www.cosmos.esa.int/gaia.  The Gaia archive
website is https://archives.esac.esa.int/gaia.  
\end{acknowledgements}

\bibliographystyle{aa}
{}

\end{document}